\documentclass[prl,reprint,showpacs]{revtex4-1}
\usepackage{amsfonts}
\usepackage{float}
\usepackage{makeidx}
\usepackage{upgreek}
\usepackage{amsmath,amssymb,graphics,epsfig,color,times,bbm}
\usepackage{ulem}
\usepackage{lineno} 

\begin{document}
	\title{High-speed quantum radio-frequency-over-light communication}
	\author{Shaocong Liang$^{1}$}
	\thanks{These authors contributed equally to this work.}
	\author{Jialin Cheng$^{1}$}
	\thanks{These authors contributed equally to this work.}
	\author{Jiliang Qin$^{1,2}$}
	\author{Jiatong Li$^{1}$}
	\author{Yi Shi$^{1}$}
	\author{Zhihui Yan$^{1,2}$}
	\email{zhyan@sxu.edu.cn}
	\author{Xiaojun Jia$^{1,2}$}
	\email{jiaxj@sxu.edu.cn}
	\author{Changde Xie$^{1,2}$}
	\author{Kunchi Peng$^{1,2}$}
	
	\affiliation{$^{1}$State Key Laboratory of Quantum Optics and Quantum Optics Devices,
		Institute of Opto-Electronics, Shanxi University, Taiyuan, 030006, P. R. China \\
		$^{2}$Collaborative Innovation Center of Extreme Optics, Shanxi University, Taiyuan 030006, P. R. China \\
	}
	\begin{abstract}
		Quantum dense coding (QDC) means to transmit two classical bits by only transferring one quantum bit, which has enabled high-capacity information transmission and strengthened system security. Continuous-variable QDC offers a promising solution to increase communication rates while achieving seamless integration with classical communication systems. Here, we propose and experimentally demonstrate a high-speed quantum radio-frequency-over-light (RFoL) communication scheme based on QDC with entangled state, and achieve a practical rate of 20 Mbps through digital modulation and RFoL communication. This scheme bridges the gap between quantum technology and real-world communication systems, which bring QDC closer to practical applications and offer prospects for further enhancement of metropolitan communication networks.
	\end{abstract}
	
	\maketitle

	Quantum physics not only promotes the emergence of a variety of novel and rapidly advancing technologies but also has the capability of combining with classic technologies, leading to the realization of extraordinary functionalities \cite{EveredSJ,GreveGP,ZuoXJ,LiuXHJ,MaLX}. The integration of quantum technology with normal communication systems offers the potential for enhancing the communication rates and security either individually or concurrently. Quantum dense coding (QDC), also known as ultradense coding, is a protocol to transmit two classical bits by transferring only one quantum bit \cite{Bennett,Nielsen,GuoY}. This is achieved when the sender and receiver share an entangled quantum pair previously, enabling a remarkable leap in data transmission efficiency \cite{Harrow,Hayden}. QDC has been successfully realized in a variety of quantum systems, including photonic qubits \cite{Mattle,Williams}, optical modes \cite{LiX,Jing}, nuclear magnetic resonance \cite{Fang}, and atom-based systems \cite{Schaetz,ZhangW}. Moreover, there is much research involved in high-dimensional QDC for increasing the communication rate \cite{HuXM,ChenY,Molina,Dada,Kues,Erhard,LiuSS} and in the establishment of multiuser QDC networks \cite{Patra,BruB,Yeo,HUXM}. The exploration of message transmission through QDC further enriches this realm \cite{Barreiro,Williams}. Within the quantum systems involved in such studies, continuous-variable (CV) QDC schemes using nonclassical states of light offer deterministic and unconditional coding, having a substantial advantage in terms of communication rate \cite{LiX,Jing,Shi,Guo}.
	
	In this Letter, we present the first demonstration of quantum radio-frequency-over-light (RFoL) communication, achieved through the synergy between QDC and classical communication systems. A 200-MHz broadband Einstein–Podolsky–Rosen (EPR) entangled state of light serves as both the quantum resource and the optical carrier. This configuration enables the transmission of digital message using the binary phase shift keying (BPSK) modulation scheme over an RF subcarrier at a practical rate of 20 Mbps. In direct comparisons with classical systems operating under identical conditions, our combined setup can offer an improved security and a markedly reduced bit error rate (BER). Due to the presence of thermal noise within a single submode of the EPR entangled state \cite{BraunsteinSL,Weedbrook}, which acts as a mask on the transmitted signal or message, thereby enhancing the system's security. Importantly, our experimental paradigm differs from previous CV-QDC methodologies, which only demonstrates the modulation and demodulation of a single-frequency signal and are far from reaching the transmission of effective and meaningful information \cite{LiX,Jing,ChenY,Shi,Guo}. In contrast, our scheme enables the high-rate transmission of digital information within the realm of QDC.
	
	Quantum RFoL communication depends on the fundamental principle of QDC in quantum optics. The use of an entangled state of light in QDC enables higher channel capacity and enhanced communication security through the introduction of quantum correlation \cite{LiX}. Specifically, for the example of an EPR entangled state with quadrature phase correlation and quadrature amplitude anticorrelation, the entangled state satisfies the following relationships $\delta^{2} (\hat{X}_{\rm EPR1}+ \hat{X}_{\rm EPR2})=\delta^2 (\hat{Y}_{\rm EPR1}- \hat{Y}_{\rm EPR2})=2 e^{-2 r}$ and $\delta^{2} (\hat{X}_{\rm EPR1}- \hat{X}_{\rm EPR2})=\delta^2 (\hat{Y}_{\rm EPR1}+ \hat{Y}_{\rm EPR2})=2 e^{2 r}$, where $r$ characterizes the degree of entanglement and the term $\delta^2 \hat{O}=\langle \hat{O}^2\rangle-\langle \hat{O}\rangle^2$ represents the variance of operator $\hat{O}$ \cite{BraunsteinSL,Weedbrook,Braunstein}. As $r \rightarrow \infty$, ideal entanglement is achieved, resulting in $\delta^{2} (\hat{X}_{\rm EPR1}+ \hat{X}_{\rm EPR2})=\delta^2 (\hat{Y}_{\rm EPR1}-\hat{Y}_{\rm EPR2})\rightarrow 0$ and $\delta^{2} (\hat{X}_{\rm EPR1}- \hat{X}_{\rm EPR2})=\delta^2 (\hat{Y}_{\rm EPR1}+ \hat{Y}_{\rm EPR2})\rightarrow \infty$. According to Shannon's theorem, the channel capacity $C$ of a Gaussian noise and Gaussian signal channel can be expressed as $C=B \log_2 (1+{\rm SNR})$, where $B$ represents the communication bandwidth and SNR is the signal-to-noise ratio. By employing one submode (EPR$_2$) of the EPR entangled state as an optical carrier to carry information and the other submode (EPR$_1$) for demodulation, a communication approach with quantum superiority can be established. Amplitude modulation and phase modulation are performed on the light field, expressed as $\hat{c}'=\hat{c}+a_{s}$, where $a_{s}$ represents the classical signal transmitted in the quantum channel. When the terms $\delta \hat{X}_c \rightarrow \delta \hat{X}_c+\delta X_{a_s}$ and $\delta \hat{Y}_c \rightarrow \delta \hat{Y}_c+\delta Y_{b_s}$ are considered, the SNRs for the modulated EPR$_2$ (as a thermal state) can be calculated as ${\rm SNR}_{X_S}=\delta^2 X_{a_s}/\delta^2 \hat{X}_c=2 \delta^2 X_{a_s}/(e^{-2 r}+e^{2 r})$ and ${\rm SNR}_{Y_S}=\delta^2 Y_{b_s}/\delta^2 \hat{Y}_c=2\delta^2 Y_{b_s}/(e^{-2 r}+e^{2 r})$. In the ideal scenario, when $r \rightarrow \infty$, ${\rm SNR}_{X_S} \rightarrow 0$ and ${\rm SNR}_{Y_S} \rightarrow 0$, implying that the signal is entirely submerged in the thermal noise. Without the necessary submode for demodulation, one cannot extract meaningful information from the modulated submode. In contrast, when EPR$_1$ is employed for joint detection of the modulated EPR$_2$, the SNRs become ${\rm SNR}_{X_S}=\delta^2 X_{a_s}/\delta^2 (\hat{X}_{\rm EPR1}+ \hat{X}_{\rm EPR2})=\delta^2 X_{a_s}/(2 e^{-2 r})$ and ${\rm SNR}_{Y_S}=\delta^2 Y_{b_s}/\delta^2 (\hat{Y}_{\rm EPR1}- \hat{Y}_{\rm EPR2})=\delta^2 Y_{b_s}/(2 e^{-2 r})$. As $r \rightarrow \infty$, ${\rm SNR}_{X_S} \rightarrow \infty$ and ${\rm SNR}_{Y_S} \rightarrow \infty$, signifying almost complete signal recovery for an authorized user with the assistance of the demodulation submode. However, infinite entanglement is not achievable in practical scenarios since it requires infinite energy. Intercepting some information with the single encoded submode of CV EPR state may be feasible, albeit with a lower SNR.
	
	\begin{figure}[htbp]
		\centering\includegraphics[width=1 \linewidth]{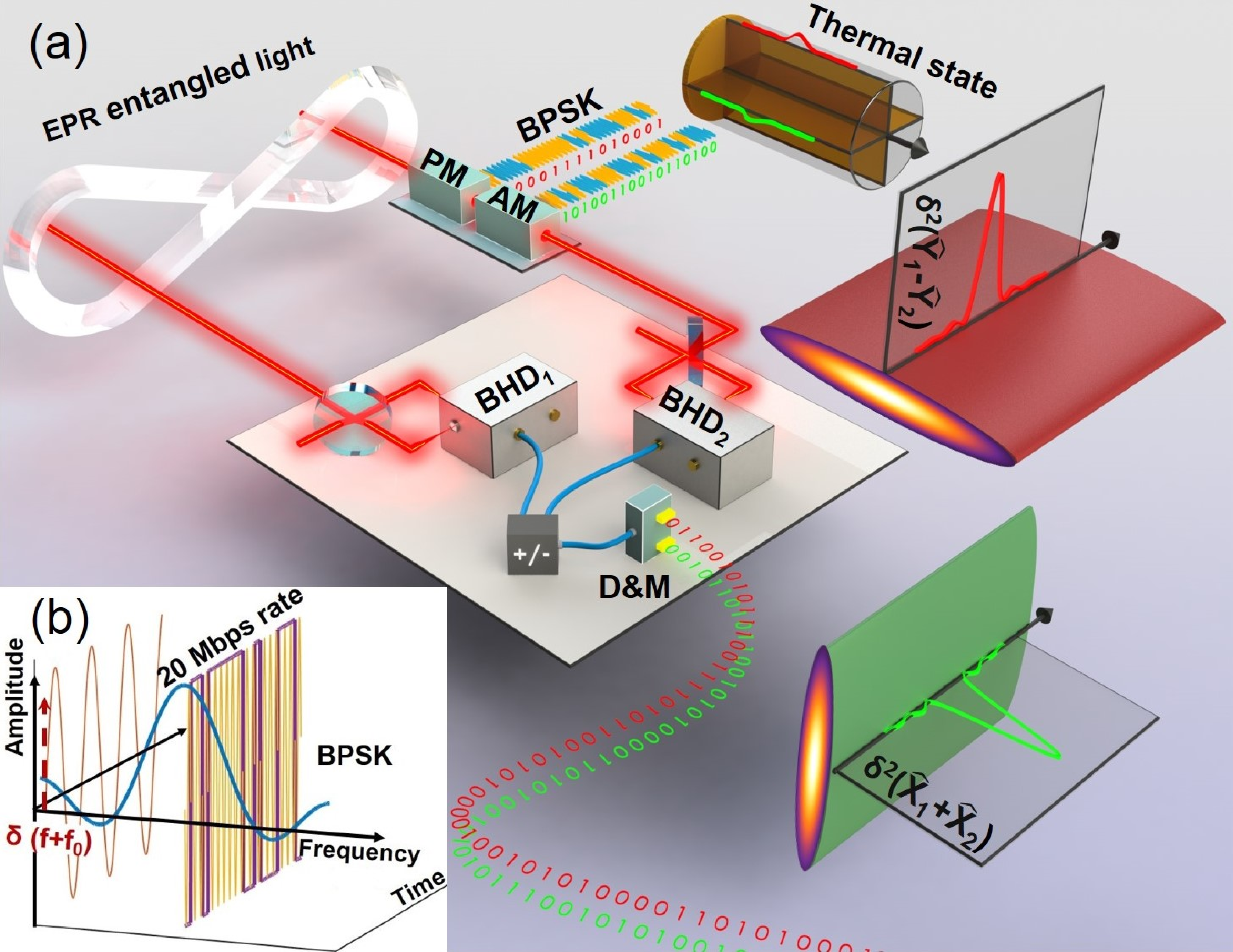}
		\caption{ \label{Fig1} Schematic diagram. (a) Schematic diagram of QDC with a broadband entangled state. The configuration involves sharing a broadband entangled state of light between Alice and Bob. Alice encodes two wideband signals onto the quadrature amplitude and phase, respectively, of the entangled submode, which is then transmitted to Bob. Bob receives the submode and decodes information with the assistance of the other submode. BPSK: binary phase shift keying modulation; AM: amplitude modulator; PM: phase modulator; BHD: balanced homodyne detector; D\&M: demodulation and measurement. (b) Comparison of the power spectra of traditional CV-QDC and our scheme. The leftmost $\delta$-shaped peak represents the traditional QDC spectrum with information typically encoded at the frequency point $f_{0}$. The right side illustrates a carrier signal capable of accommodating high-speed digital information, characterized by a specific frequency bandwidth.}
	\end{figure}
	
	In the proposed system, the main strategy for increasing the communication rate is to improve both the SNR and the communication bandwidth. Notably, the expansion of the squeezed state's bandwidth, as previously demonstrated in Refs. \cite{Ast,Kashiwazaki1,Kashiwazaki2,Inoue,Kashiwazaki3}, leads to highly promising results in this regard. Moreover, the chosen encoding method also influences the achievable communication rate. Given its robust noise resistance, straightforward methods for evaluating communication quality, and compatibility with existing communication systems, the digital modulation method known as BPSK is opted to encode and decode information. This is significantly different from the traditional QDC schemes. The communication system is capable of transmitting effective digital information at a communication rate comparable with practical applications when a broadband nonclassical optical field is used as the carrier. This is determined by the inherent nature of a digital signal, which occupies specific communication frequency bands, as depicted in Fig. \ref{Fig1}b. Therefore, we exploit the QDC method based on a broadband entangled state of light, which serves as the foundation for the quantum RFoL communication scheme. The fundamental framework of our approach is illustrated in Fig. \ref{Fig1}a. Sharing a pair of entangled submodes between the sender Alice and the receiver Bob is a prerequisite for the communication. Alice encodes her submode (thermal state) and subsequently distributes it to Bob. The submode received by Bob is demodulated with the assistance of Bob’s submode, thereby recovering the information. Crucially, our experimental configuration introduces an RF signal as a subcarrier of the digital information. This subcarrier is further modulated onto the entangled submode. Digital Information can be transmitted in fiber or free space using this method, and the resulting architecture supports subsequent information demodulation.
	
	First, we construct a classical RFoL communication system grounded in a coherent state of light, in which the information received by Bob is the result of classical communication (see Supplemental Material \cite{Supple}). Then, a 200-MHz EPR entangled state of light is subsequently introduced into the classical communication framework to achieve the quantum RFoL communication system based on the QDC method. A pair of 10-mm semi-monolithic optical parametric amplifiers (OPAs) contain a 6-mm periodically poled $\rm KTi OPO_{4}$ crystal and a concave mirror with a radius of curvature of 50 mm. When these two OPAs operate below their oscillation threshold, two amplitude-squeezed states with the same frequency and spatial mode are prepared. The squeezing levels of the squeezed state Sq1 (Sq2) are $7.5 \pm 0.3$ dB ($7.0 \pm 0.2$ dB), $5.9 \pm 0.3$ dB ($5.7 \pm 0.2$ dB) and $2.2 \pm 0.2$ dB ($2.1 \pm 0.2$ dB) below the shot noise limit (SNL) at 3 MHz, 63 MHz and 200 MHz, respectively. Subsequently, a 200-MHz EPR entangled light is obtained by combining these two squeezed states onto a 50/50 beam splitter with a phase difference of $\pi/2$ \cite{BraunsteinSL,aFurusawa}. The correlation variances of the amplitude sum (phase difference) are $6.4 \pm 0.2$ dB ($6.4 \pm 0.3$ dB) at 3 MHz and $2.0 \pm 0.1$ dB ($1.9 \pm 0.1$ dB) at 200 MHz below the corresponding SNL, respectively. The correlation variance in frequency range 23 MHz to 63 MHz is relatively equilibrated. The nonclassical states are measured using a homemade balanced homodyne detector characterized by its low noise, high SNR, and relatively flat response within a 200 MHz bandwidth. This detector is composed of two photodiodes FD100 (Fermionics Opto-Technology) with a junction capacitance of 1.1 pF, two operational amplifiers OPA855 (Texas Instruments) and corresponding two-stage amplifier configurations (see Supplemental Material \cite{Supple} for details). This entangled state is prepared and provided by Bob, the designated receiver in the experiment. Bob reserves one submode (EPR$_1$) for the recover of information from Alice and simultaneously distributes the second submode (EPR$_2$) to Alice. Alice modulates the classical information on the received EPR$_2$ and then sends EPR$_2$ back to Bob. Finally, Bob executes joint balanced homodyne detection with the assistance of EPR$_1$ and recovers the classical information. This scheme incorporates a round-trip transmission protocol, thereby protecting the demodulation beam EPR$_1$ from potential exposure. This defensive measure effectively enhances the security of the system, preventing interception and forwarding attacks.
	
	\begin{figure}[ht]
		\centering\includegraphics[width=0.85 \linewidth]{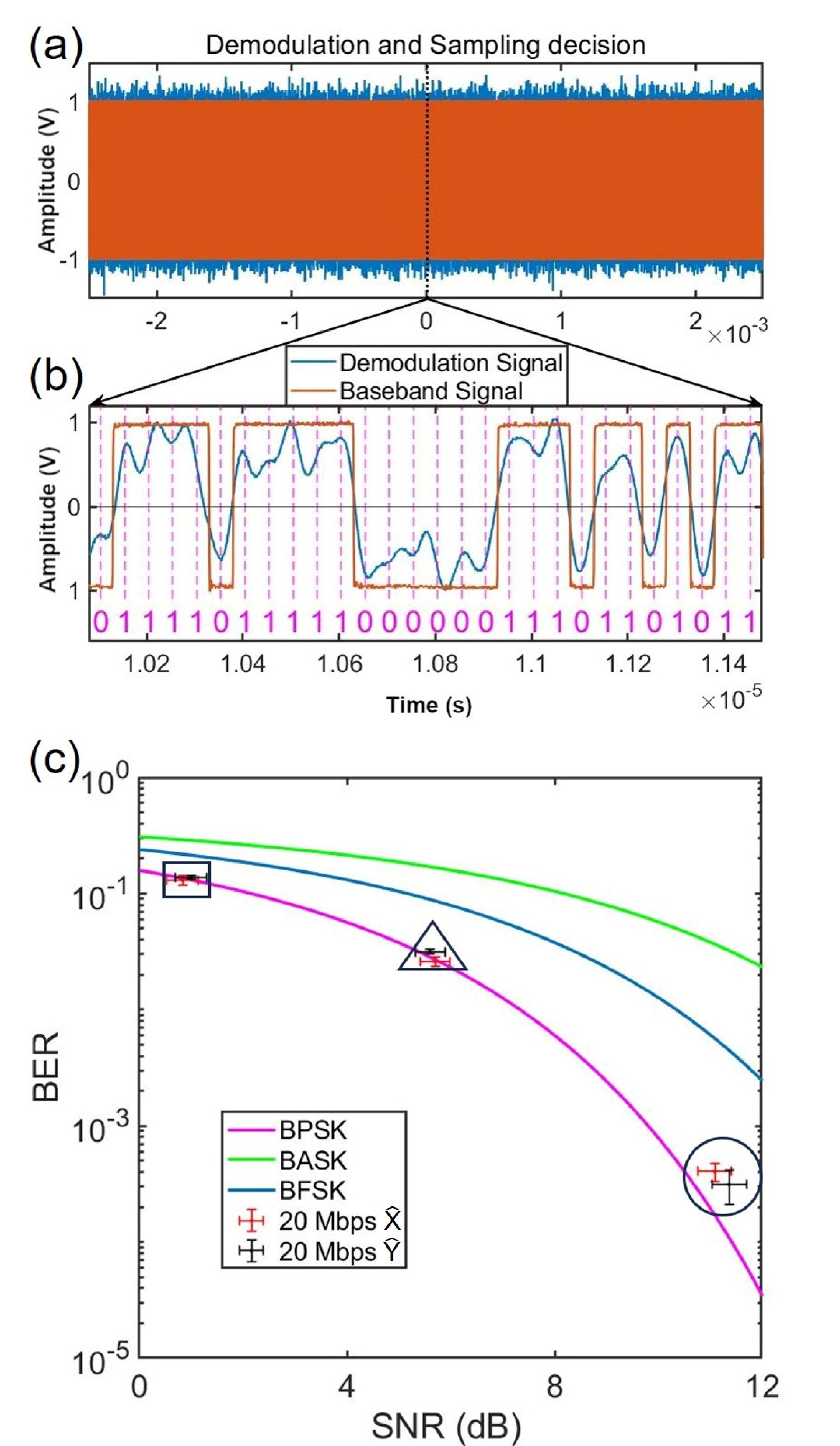}
		\caption{ \label{Fig3} Experimental results. (a) Comparison between the baseband signal and the demodulated signal on the oscilloscope. The orange and blue curves are the original signal and the demodulated signal with quadrature phase correlation, respectively. (b) An enlarged view of the period marked by the dotted line in (a) (from 10.1 $\upmu$s to 11.5 $\upmu$s) along with the symbol decision results. (c) BER comparison. The green, blue, and magenta curves represent the relationships between the BER of the demodulated information and the SNR achieved with the BASK, BFSK and BPSK modulation methods, respectively. The red and black dots indicate the actual BERs obtained in the experiment. These values are determined via quadratures encoding and decoding using the BPSK modulation technique. The points within the square, triangular and circular regions represent the results measured with the single submode (thermal state), the coherent state and the combined submodes (entangled state), respectively, where the same modulation signal is used in all three cases. The error bars are derived from 5 consecutive BER measurements each based on $10^{5}$ bits.}
	\end{figure}
	
	Subsequently, we demonstrate the transmission and receipt of real messages by means of the quantum RFoL system (see Supplemental Material \cite{Supple} for detailed experimental setup). Constrained by the frequency response characteristics of the phase modulator (M4004, New Focus) and the entanglement characteristics of the EPR entangled light, we select a frequency band ranging from 23 to 63 MHz for information transmission while maintaining a fixed transmission rate of 20 Mbps. The BPSK method is employed to modulate digital information onto the RF signal, and the modulated RF signal is then loaded onto the optical carrier, namely, the entangled submode EPR$_2$. In this process, the digital signal is generated from a baseband signal source (BBSG). The actual transmitted digital message takes the form of a string of bipolar non-return-to-zero codes. After the RF signal is loaded with a baseband signal, it is used to modulate the amplitude and phase of the submode EPR$_2$. The sender Alice transmits EPR$_2$ modulated with the RF signals carrying information to the receiver Bob. Bob applies EPR$_1$ and the received EPR$_2$ to joint balanced homodyne detection. With the assistance of classical memory, Bob is able to demodulate the information from the quadrature amplitude sum and phase difference of the pair of entangled submodes. Access to the separate submode EPR$_2$ is limited to the thermal state, where only signals with low SNR can be obtained. This is achieved by demodulating EPR$_2$ using a single homodyne detector instead of joint homodyne detection in quantum scenario. Based on the same digital message for transmission, the power spectra as measured with the coherent state (the classical counterpart), single submode and combined submodes have been obtained (see Supplemental Material \cite{Supple}). The information decoded in the quadrature phase and amplitude within the QDC communication system, supported by the shared entangled submodes, exhibits an enhancement in SNR, $5.7 \pm 0.5$ dB for amplitude and $5.6 \pm 0.4$ dB for phase relative to the classical communication system. Conversely, measuring single submode only results in a decline in SNR, $4.6 \pm 0.3$ dB for amplitude and $4.9 \pm 0.3$ dB for phase.
	
	In an assessment analogous to that for traditional digital communication systems, the BER of the quantum RFoL communication is a pivotal parameter. While fixing the transmission rate of digital information at a consistent 20 Mbps, the BERs of different systems are measured and evaluated experimentally. Using the submode EPR$_2$, Bob performs joint balanced homodyne detection to obtain RF subcarrier carrying information, and performs coherent demodulation on RF signal to recover baseband signal. This restored baseband signal is fed to an oscilloscope (OSC) for acquisition. Bit synchronization, sampling and symbol decision are carried out to obtain the information received by Bob. By comparing this information with the original baseband signal sent from Alice, the experimental BERs for amplitude and phase modulation are estimated quantitatively. In Fig. \ref{Fig3}a, the baseband signal obtained after transmission and the original baseband signal are illustrated visually. Figure \ref{Fig3}b provides an enlarged view of the interval from 10.1 $\upmu$s to 11.5 $\upmu$s in Fig. \ref{Fig3}a. Moreover, our evaluation involves a theoretical comparison of the BERs achieved with three fundamental digital modulation techniques: BPSK, binary frequency shift keying (BFSK) and binary amplitude shift keying (BASK), all employing identical modulation signals (see Supplemental Material \cite{Supple} for the theoretical results). As shown in Fig. \ref{Fig3}c, BPSK modulation distinctly excels in terms of anti-noise performance. Figure \ref{Fig3}c also includes the BERs achieved with entangled state, thermal state and coherent state in the RFoL communication system, which are used to represent the communication quality using joint homodyne detection, single submode, and the corresponding classical system, respectively. According to Fig. \ref{Fig3}c, when information is encoded onto the quadrature amplitude (phase), the receiving end incurs a BER of $(2.6 \pm 0.2)\times10^{-2}$ (($3.2 \pm 0.2)\times10^{-2}$) in classical communication. For a quantum user, the exploitation of entangled state leads to a remarkable reduction in the BER, reaching $(4.0 \pm 0.7)\times10^{-4}$ for amplitude encoding and $(3.0 \pm 0.9)\times10^{-4}$ for phase encoding. In contrast, the BER is magnified to $(1.3 \pm 0.1)\times10^{-1}$ for amplitude encoding and $(1.4 \pm 0.1)\times10^{-1}$ for phase encoding without the demodulation submode. It is clear that the adoption of entangled state brings significant advantages, reducing the BER by two orders of magnitude by using joint homodyne detection and increasing the BER by three orders of magnitude when only the single submode is used. Therefore, QDC exploiting broadband entangled submodes leads to a reduced BER and improved security. The former implies a quantum-enhanced improvement in communication rate.
	
	\begin{figure}[htbp]
		\centering\includegraphics[width= 0.98\linewidth]{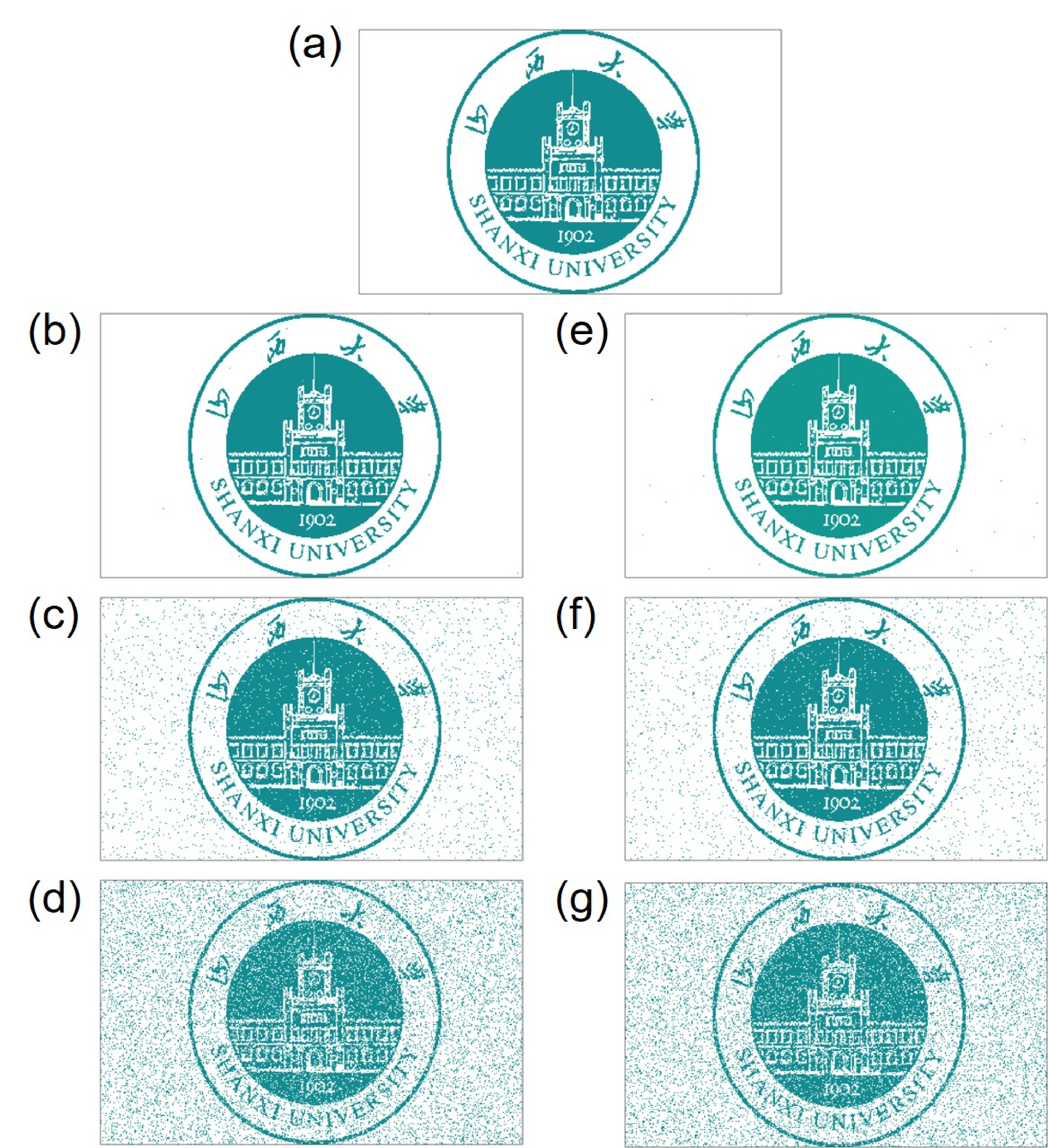}
		\caption{ \label{Fig4} Comparison between the original and received images when using different communication methods. (a) The original dichroic $250\times 400$ pixel image to be transmitted. (b)-(d) ((e)-(g)) Images obtained by performing encoding and decoding on the quadrature phase $\hat{Y}$ (amplitude $\hat{X}$) of the optical carrier, where (b) ((e)), (c) ((f)) and (d) ((g)) are the images received with entangled state, coherent state and the single submode, respectively.}
	\end{figure}
	
	Furthermore, a dichroic image is employed as a concrete binary message to be transmitted within our quantum framework to visually demonstrate the advantages over the corresponding classical system. We compare the received images across three different scenarios: classical communication, using the single submode and entangled submodes in quantum communication, as shown in Fig. \ref{Fig4}. The image obtained via quantum RFoL communication exhibits better clarity compared to its classical counterpart. The image quality achieved through classical communication is higher than that of the image without demodulation submode in the quantum scenario. These results not only validate the efficacy of our proposed approach but also visually demonstrate its significant advantages. Moreover, these results emphasize the quantum security aspect, demonstrating that intercepted image using only the encoded submode in quantum settings are markedly less clear than its classical counterpart. As an extension of our work, a succinct and straightforward protocol for the quantum RFoL communication based on QDC method is outlined (see Supplemental Material \cite{Supple}), providing a simple framework for practical implementation and further development.
	
	In conclusion, we realize a quantum RFoL communication by employing QDC at a transmission rate of 20 Mbps on both two quadratures. With a practical BER of $10^{-4}$ prior to error correction, our system not only meets but exceeds the minimal BER standard required for commercial optical communication, which typically demands a BER not surpassing $10^{-3}$. This achievement firmly establishes the feasibility and potential of quantum RFoL communication. On the other hand, the quantum RFoL system effectively conceals partial information due to the inherent thermal-noise properties of the EPR submode. By seamlessly integrating quantum technology with classical communication systems, the proposed quantum RFoL scheme significantly expands the potential applicability of QDC. Our experiment represents a pivotal stride towards the widespread adoption of QDC in conventional communication systems.
	
	Despite providing a significantly enhanced communication rate compared to previous QDC systems, our communication rate remains lower than that of classical RFoL communication due to the lower frequency of the RF carrier \cite{ZhangCB}. This limitation arises from the constraints imposed by the bandwidths of the entanglement source and RF devices. Excitingly, recent advancements in waveguide-based ultrabroadband squeezed light technology, featuring an ultrawide bandwidth of the squeezed state, hold promise for overcoming these constraints \cite{Kashiwazaki1,Kashiwazaki2,Inoue,Kashiwazaki3}. Exploiting these breakthrough technologies, quantum RFoL communication can potentially achieve rates equivalent to or even surpassing those of commercial telecommunication. Such progress will carry profound implications for the practical implementation of quantum information technology.
	
	This work was supported by the National Natural Science Foundation of China (Grants No. 61925503, No. 62122044 and No. 62135008), the Program for the Innovative Talents of the Higher Education Institutions of Shanxi, the Program for the Outstanding Innovative Teams of the Higher Learning Institutions of Shanxi and the Fund for Shanxi \textquotedblleft 1331 Project\textquotedblright\ Key Subjects Construction.

\end{document}